\PassOptionsToPackage{unicode}{hyperref}
\PassOptionsToPackage{naturalnames}{hyperref}

\documentclass[times,twocolumn,final,authoryear]{elsarticle}
\usepackage{jasr}
\usepackage{framed,multirow}

\usepackage{amssymb}
\usepackage{latexsym}

\newcommand{\kms}{km s$^{-1}$}

\usepackage{url}
\usepackage{xcolor}
\definecolor{newcolor}{rgb}{.8,.349,.1}

\usepackage[
  breaklinks = true,
   colorlinks = true,
  urlcolor   = blue,
  citecolor  = blue,
  linkcolor  = blue,
            pdfauthor=author,
]{hyperref}
\journal{Advances in Space Research}

\begin{document}

\pdfstringdefDisableCommands{%
  \def\\{}%
  \def\texttt#1{<#1>}%
}

\verso{Brigitte Schmieder, Reetika Joshi and Ramesh Chandra}%\textit{etal}}

\begin{frontmatter}

\title{Solar jets observed with the Interface Region Imaging Spectrograph (IRIS)}

\author[1,2,3]{Brigitte \snm{Schmieder}\corref{cor1}}
\ead{Brigitte.Schmieder@obspm.fr}
\cortext[cor1]{Corresponding author 
 }
\author[4]{Reetika \snm{Joshi}}
\ead{reetikajoshi.ntl@gmail.com}
%\fnref{fn1}}
%\fntext[fn1]{This is author footnote for second author.}
\author[4]{Ramesh \snm{Chandra}}
\ead{rchandra.ntl@gmail.com}

%% Third author's email
%\ead{}
%\author[2]{Given-name4 \snm{Surname4}}

\address[1]{LESIA, Observatoire de Paris, Université PSL, CNRS, Sorbonne Université, Université de Paris, 5 place Jules Janssen, 92190 Meudon, France}
\address[2]{Centre for Mathematical Plasma Astrophysics, Dept. of Mathematics, KU Leuven, 3001 Leuven, Belgium}
\address[3]{University of Glasgow, Glasgow, Scotland}
\address[4]{Department of Physics, DSB Campus, Kumaun University, Nainital -- 263 001, India}
%\received{1 May 2013}
%\finalform{10 May 2013}
%\accepted{13 May 2013}
%\availableonline{15 May 2013}
\communicated{B. Schmieder}

\begin{abstract}
{Solar jets are  impulsive, collimated  plasma   ejections  that are triggered by magnetic reconnection.}
{They are observed for many decades in various temperatures and wavelengths, therefore 
their  kinematic characteristics, such as  velocity and recurrence, have been extensively studied.}
Nevertheless, the high spatial resolution of the Interface Region Imaging Spectrograph (IRIS) launched in 2013  allowed us to make a step  
{forward in the understanding of the relationship} between surges and hot jets. In this paper we report on several results  of recent 
{studies of jets observed by IRIS}. Cool and hot plasma  have been detected with ejections  of cool blobs having a speed reaching 300 km s$^{-1}$ during  the impulsive phase of  jet formation and slow velocity surges surrounding  hot jets after the reconnection phase.
{Plasma characteristics of solar jets, such as the emission measure, temperature, and density have been quantified.}
A multi-layer atmosphere  at the reconnection site based on observed  IRIS spectra has been proposed. 
IRIS  evidenced bidirectional flows at  reconnection sites,  and  tilt along the spectra which were  interpreted as   the signature of twist in jets.
 The search of possible sites for  reconnection could be achieved by  the analysis of magnetic topology.
Combining {Solar Dynamics Observatory/Helioseismic Magnetic Imager (SDO/HMI)} vector magnetograms and  IRIS observations, it was found that 
reconnection site could be located at  null points in the corona as well as in bald patch regions low in the photosphere.
In one case study a   magnetic sketch could explain   the  initiation of  a jet starting in a bald patch transformed  to a current sheet in a dynamical way, and   the  transfer of   twist
from  a flux rope 
to the jet during the magnetic reconnection  process.
\end{abstract}

\end{frontmatter}

\section{Introduction}
\label{sec:intro}
Jets are plasma ejections along collimated magnetic field lines. They 
act as a source for transporting mass and energy and
{can contribute to the heating of the solar corona and the acceleration of the solar wind.}
They are the key tool to probe the broad dimensions of the heliospheric problems.
{Many authors have analysed these small-scale solar ejections, from}
H$\alpha$ surges \citep{Roy1973,Schmieder1983, Uddin2012} to X-ray jets \citep{Shibata1992} or X-ray coronal loops \citep{Schmieder1995}. 
All the literature concerning jets from a theoretical point of view as well from observations is summarized in recent reviews \citep{Raouafi2016,Pariat2016,Shen2021,DePontieu2021}. 

Several  recent {studies focused on recurrent homologous jets observed in different}
wavelengths such as in H$\alpha$ \citep{Asai2001,Chandra2017}, in EUV \citep{Chae1999, Jiang2007,Schmieder2013, Chandra2015, Joshi2017b}, and in X-rays \citep{Kim2001,Kamio2007,Sterling2016}.
The characteristics of jets {are} very well described from  jet  observations made  by the Atmospheric Imaging
Assembly (AIA, \citet{Lemen2012}) telescope on board the Solar Dynamics Observatory (SDO, \citet{Pesnell2012}) satellite  \citep{Kayshap2017,Joshi2017b}.  The range of jet velocities is between 100 to 300 \kms and the length {can be as short as 5 Mm and as long as 100 Mm.}
\begin{figure*}[ht!]
\centering
\includegraphics[width=\textwidth]{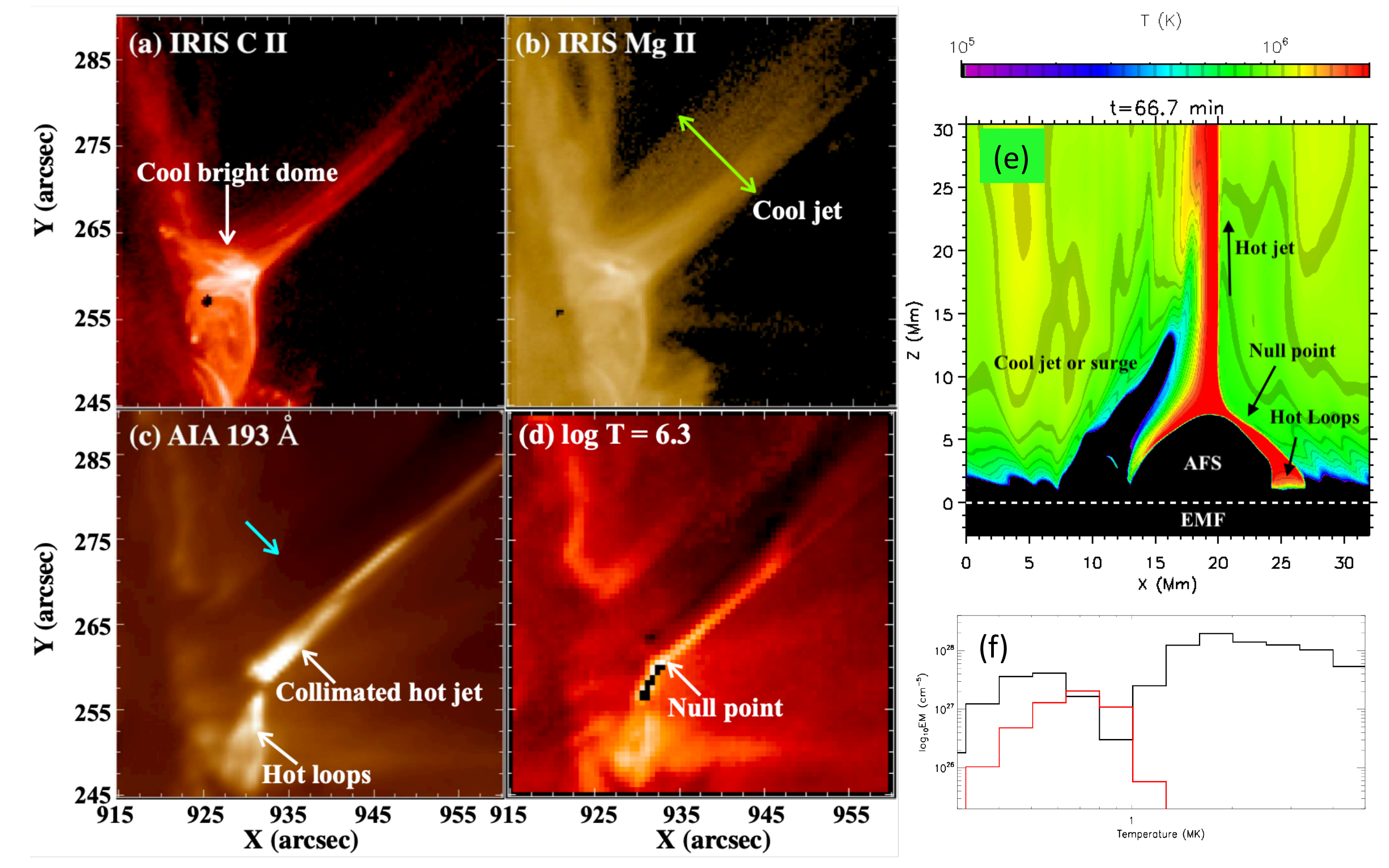}
\caption{IRIS slit jaw images  of a jet on April 4 2017 observed in cool plasma (C\textsc{ii} and Mg\textsc{ii}) (top panels a-b) and  in hot plasma in  AIA 193 \AA\ image and temperature map for log T (K) = 6.3 (panels c-d).
 Panel (e): numerical simulation of a cool surge surrounding a hot jet during reconnection  fitting well with the jet observations. Panel (f):  emission measure (EM) in the jet (red) and its base (black) (adapted from  \citet{Nobrega2017,Joshi2020}).}
\label{Fig_Nobrega}
\end{figure*}
Trigger mechanisms of jets  are   based on  magnetic reconnection depending on the magnetic configuration of the region  with 
flux emergence and cancellation \citep{Archontis2009,Torok2009,Moreno2013,Priest2018} or with  rotation of sunspot \citep{Pariat2009b,Curdt2012}. Jets occur at  different locations,  in  the center or at the border of active region \citep{Chandra2017,Joshi2017b}, at the edge of  coronal holes \citep{Moreno2008,Moreno2013} or {at } light-bridges \citep{Robustini2016,Yang2020,Bharti2020}. {Spicules are also considered as chromospheric jets  and well studied with recent instrumentation \citep{DePontieu2021}.} {A new kind of jet/spicule has been discovered with IRIS related to the activity of filaments and prominences. Photospheric motions are favouring the braiding of magnetic field lines which could interact between them  and produce nano jets  through reconnection in loops and prominences \citep{Chen2017,Huang2018,Chitta2019,Chen2020,Antolin2021}.}
{The motivation of this paper is based on new imaging and spectral observations obtained with IRIS which covers 
a large  chromospheric and transition region temperature range. Therefore  they are very complementary to  the 
high spatial and temporal imaging observations of  SDO/AIA instrument. We present a few studies of jets observed with IRIS in a broader context \citep{Joshi2017b,Joshi2020,Joshi2020FR,Joshi2021b,Joshi2021a}.}
{These observations allow us  to answer several questions. What are the plasma characteristics of jets (temperature, density)? Section \ref{sec3.1} is focused on hot and cool plasma properties. 
Spectral diagnostics with O\textsc{iv} lines  allow to   retrieve  accurate density values for different types of jets.  The spire  jet density   is   one or two orders of magnitude less than the  jet base  density.
What is the dynamic of jets in 3D?
Section \ref{sec3.2} presents the explosive dynamic nature {of jets, their}  morphology  with examples of bilateral flows, rotation and twist measured  in  IRIS spectra.
Why is there cool material over hot material in the flare site? 
 From where the twist comes in the jets?
  proposed  during magnetic reconnection at the jet base from a flux rope to a jet  with plasmoid ejections is presented (Sections \ref{sec4.1} and \ref{sec4.2}). 
  We discuss  on a multi thermal flare model deduced from the IRIS spectra.
  Electric currents measured  at the jet base suggest the formation of quasi-separatrix  layers (Section \ref{sec4.3}).
Finally we  discuss on the benefits from the observations of IRIS in relatively cool temperature jets
 for braiding loops and nano jets  (Section \ref{sec4.4}).}
\section{Observations}
\label{sec:obs}
IRIS \citep{Pontieu2014} observed many jets since its launch in 2013 in different  modes of observations, from very dense,  medium or  coarse rasters and, simultaneously with slit jaw images (SJI)  having the same coordinate center  with a  maximum field of view of 167 arcsec  $\times$ 175 arcsec. Usually IRIS observes in four channels around  1330 \AA,  1400 \AA, 2796 \AA, 2832 \AA\ including C\textsc{ii}, Si\textsc{iv}, Mg\textsc{ii} lines, and the UV continuum, respectively. 
C\textsc{ii} lines are formed around T = 30000 K and Si\textsc{iv} lines around 80000 K, while Mg\textsc{ii} lines are  formed at chromospheric temperatures between 8000 K and 20000 K. The cadence of  SJIs is  a few seconds to one minute  and the pixel resolution size is 0.35 arcsec. 
The field of view of the rasters can be reduced to one slit (sit and stare mode) or 4 slits (field of view 16 arcsec $\times$ 60 arcsec) or 32 slits (FOV = 62 $\times$ 60 arcsec). From the rasters Dopplershift line width  can be directly retrieved for  each pixel after  fitting the profiles with Gaussian function  {routine available in Solarsoft.}
 IRIS  brings to us an important knowledge about chromospheric and transition region plasma and complement AIA images which allow to compute with accuracy the transverse velocity  of jets.
%fig2
\begin{figure*}[ht!]
\centering
\includegraphics[width=\textwidth]{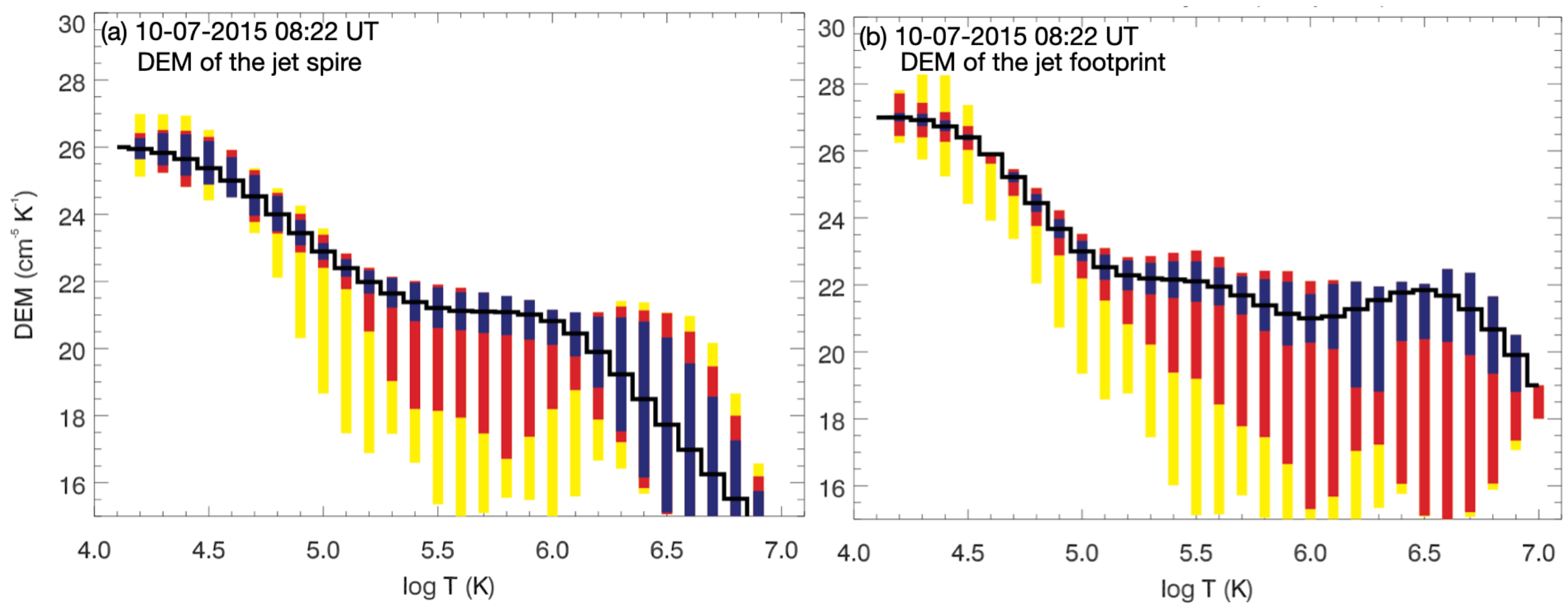}
\caption{Differential emission measure (DEM) obtained by combining cool lines of IRIS and AIA filter data of July 10 2015. The  estimates of
uncertainties on the best-fit DEM curves  are obtained by randomly varying the
input intensities by 20\% after  performing 300 Monte Carlo (MC)
simulations on the intensities. In each temperature bin, the errors
are plotted with different colour bars. 50\% of solutions closest to
best-fit DEM are shown by blue, 80\% by red and 95\% by yellow. (adapted from \citet{Mulay2017DEM}.
}
\label{Fig_DEM}
\end{figure*}
\section{Solar jets observed with IRIS}
\label{sec3}
%\subsection{Hot and cool jet - plasma characteristics}
\subsection{Plasma characteristics of hot and cool jets}
\label{sec3.1}
{%IRIS showed in many cases the presence of cool plasma during the ejection of hot plasma.  
IRIS with its spectral mode allowed us to get quantitative results on the emission measure (EM), temperature and electron density.} 

{Working on 
six recurrent jets  in the active region NOAA 12644 on April 4, 2017,  cool IRIS surges as well as hot AIA filter jets were detected by \citet{Joshi2020} (Fig. \ref{Fig_Nobrega}).} 
A fine co-alignment of the AIA and IRIS data shows that 
the hot jets are collimated  and observed in the hot temperature filters. They  have
high velocities, around 250 km s$^{-1}$ and are accompanied by cool surges surrounding the hot jets and ejected kernels that both move at about 45 km s$^{-1}$. 
The jets were initiated at  a null point at the top of a canopy-like double-chambered
structure with cool emission on one  side and hot emission on the other side as proposed {in the} emerging flux simulations of \citet{Nobrega2017}. 
{\citet{Joshi2020} used the AIA filters to study}
%In \citet{Joshi2020}  the AIA filters allowed them to study 
the temperature and the %differential 
EM of the jets using the filter ratio method {(Fig. \ref{Fig_Nobrega} panel (f))}. {The EM (cm$^{-5}$) reaches 10$^{28}$ for coronal temperature in the jet base and 10$^{27}$ in the jet spire for this case.}

Cool and hot emission in  other recurring active region jets with
significant cool emission at   foopoints  and along the jets are frequently observed  with IRIS \citep{Mulay2016,Mulay2017,Mulay2017DEM,Kayshap2021,Nobrega2021}.
%Three order of magnitude higher for the 
 %y  taking into account of IRIS  data to compute DEM
%In their  DEM analysis 
 The DEM analysis in \citet{Mulay2017DEM} took into account the IRIS data  with the EUV  AIA data of  the  cool  and hot emission  in the spire and the footpoint regions (Fig. \ref{Fig_DEM}). The  emission was peaked
at log T (K) = 5.6–5.9 which is a transition zone temperature  and 6.5 which is coronal emission, respectively. The differential emission measure (DEM) curves show the presence of hot plasma (T = 3 MK) in the footpoint region. They confirmed this result by estimating the Fe XVIII emission from the AIA 94 \AA\ channel which was formed at an effective temperature
of log T (K) = 6.5. The average 
EM (cm$^{-5}$) was increasing by three orders of magnitude reaching 10$^{31}$ for the spire  and 10$^{32}$  for the jet  base compared with results using only AIA data  \citep{Mulay2017,Mulay2018,Joshi2020,Kayshap2021}.
%(Fig. \ref{Fig_DEM}).
 
\begin{figure}
\centering
\includegraphics[width=0.48\textwidth]{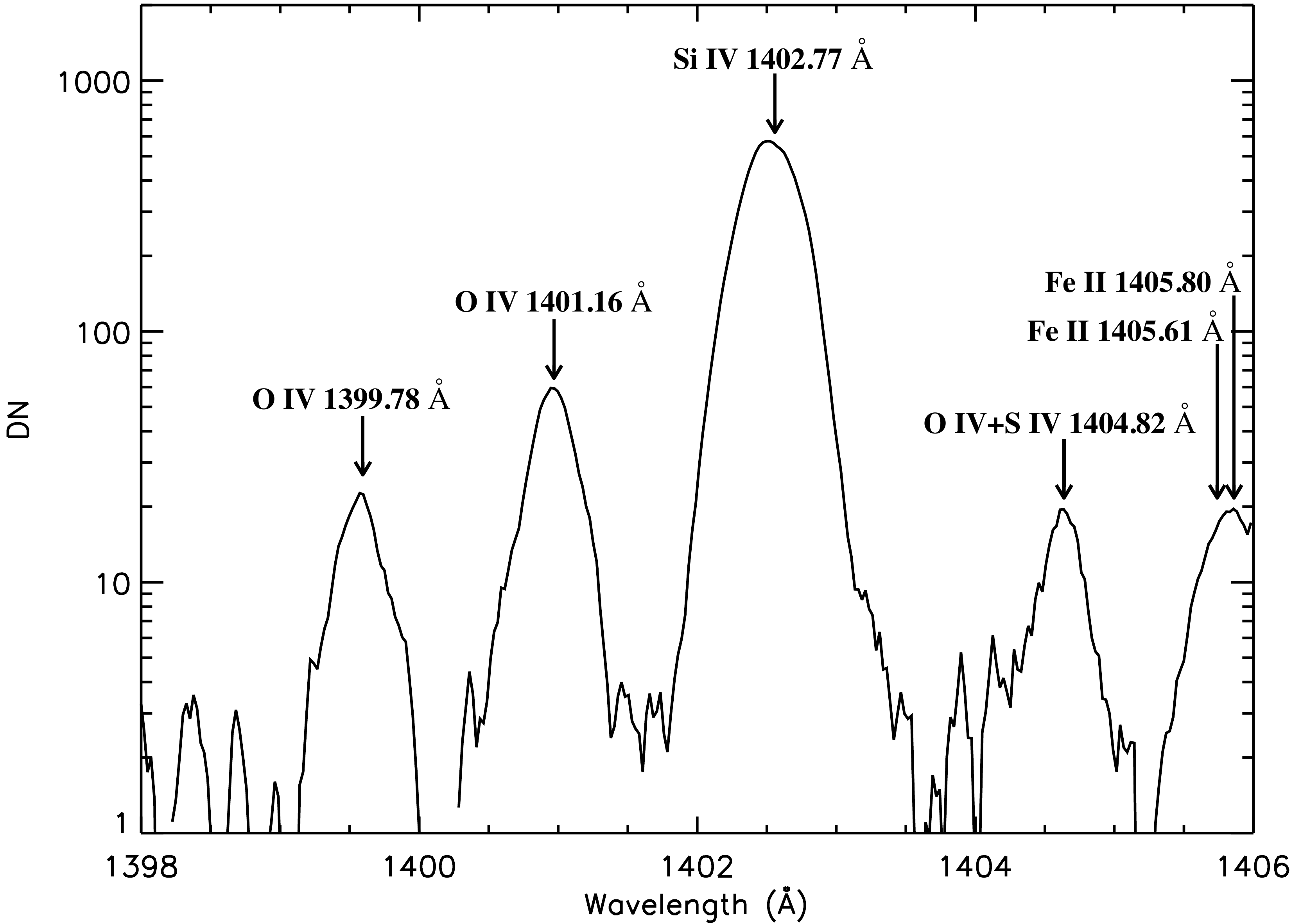}
\caption{IRIS spectra  (in log DN units) for the footpoint region of jet obtained for the Si\textsc{iv} 1403 \AA\ window  on March 22 2019 \citep{Joshi2021a}. At the time of the reconnection O\textsc{iv} lines are detectable in the mini flare.}
\label{OIV}
\end{figure}

%SPECTRA of IRIS
%IRIS Density from O IV ratio
In the  IRIS spectra  O\textsc{iv} lines are observed in the Si\textsc{iv} waveband at the location of jets (Fig. \ref{OIV}).
%(figure of Si IV and OIV spectra)?
O\textsc{iv} lines observed in the IRIS FUV spectra allow  to perform density diagnostics on the jet material  at transition region temperatures. 
The ratio of 
O\textsc{iv}
%(1399/1401 \AA) 
intensities allow to compute the electron densities  in case of Si\textsc{iv} is not too optically thick \citep{Judge2015}. \citet{Cheung2015} used two diﬀerent line ratios to measure the densities. The ﬁrst is the ratio of the O\textsc{iv} {1401.16 and 1404.82} lines in jets. The {1404.82} line is blended with a S\textsc{iv} line, so they used the S\textsc{iv} 1406.02 line  assuming  an optically thin emission to extract the intensity of O\textsc{iv} 1404.8. The O\textsc{iv} 1404.8 line is not always present in the spectral readout window of IRIS. However, in a number of slit positions where there is suﬃcient blueshift, they  found the ratio O\textsc{iv} 1401.1/1404.8 to be in the range 4.0-4.5. For a temperature range of {log T (K)} = 4.5-5.5 (derived from the ratio of S\textsc{iv} 1404.8 to S\textsc{iv} 1406.02), this ratio gives densities ranging from log Ne (cm$^{-3}$) = 10.8-11.0. Similarly, a ratio computed for the 1399.77 and 1401.16 lines of O\textsc{iv} has values in the range 0.29-0.35, which yields  electron densities of log Ne (cm$^{-3}$) = 10.8-11.2.

{Using the same diagnostics  rather low values for  temperature  (log T (K) = 5.42)  and  density  (log Ne (cm$^{-3}$) = 10.3)  were derived  for short  corona  mini-jet in clusters above  activated prominence  fine structures \citep{Chen2020}.}

\citet{Cai2019} found 
the temperature of the plasma in the jet  heated to atleast T= 10$^{5.6}$ K. {The electron density  is about log Ne (cm$^{-3}$) = 11 according to the intensity ratios of} the {O\textsc{iv} 1399.77/1401.16} doublet and {Si\textsc{iv} 1402.77/O\textsc{iv} 1401.16} lines. 
%In the vicinity of Si IV lines, OIV lines are detected  in brightarea, the base of the jet and eventually in the jet.
 \citet{Mulay2017DEM} found in their analysis electron {densities (cm$^{-3}$)} %{\rc values} 
 around  10.3 
 %2.0 $\times$ 10$^{10}$
 in the spire and 10.9  
 %7.6 $\times$ 10$^{10}$ 
 in the footpoint while \citet{Joshi2021a}  found in the jet canopy
%the resulting densities are 
roughly log Ne (cm$^{-3}$) = 11 $\pm$ 0.3.  At  the  base of  jet where the reconnection occurred  the  shape of the profiles did not allow  them to distinguish if there were many components or if it was due to microturbulence, therefore the electron density diagnostics was difficult to be  applied \citep{Joshi2021a}.
 
 The study by \citet{Nobrega2021}  concerns O\textsc{iv} emitting layers around  cool surges (T=6000 K) observed  in  Mg\textsc{ii} line profiles. With the  O\textsc{iv} ratio diagnostics they found electron density (cm$^{-3}$) values of the order of {1.6 $\times$  10$^{11}$ to 10$^{12}$.} This result was  confirmed by  numerical experiments performed with the Bifrost code.
%very low velocities. 
Those higher values for  jets 
 correspond to those obtained  for H$\alpha$ surge \citep{schmieder1988,Schmieder1995}  and   for transition regions like   in  plage and  bright point \citep{Polito2016}.

%fig3

%fig4
\begin{figure*}[ht!]
\centering
%\vspace{1.5cm}
\includegraphics[width=0.9\textwidth]{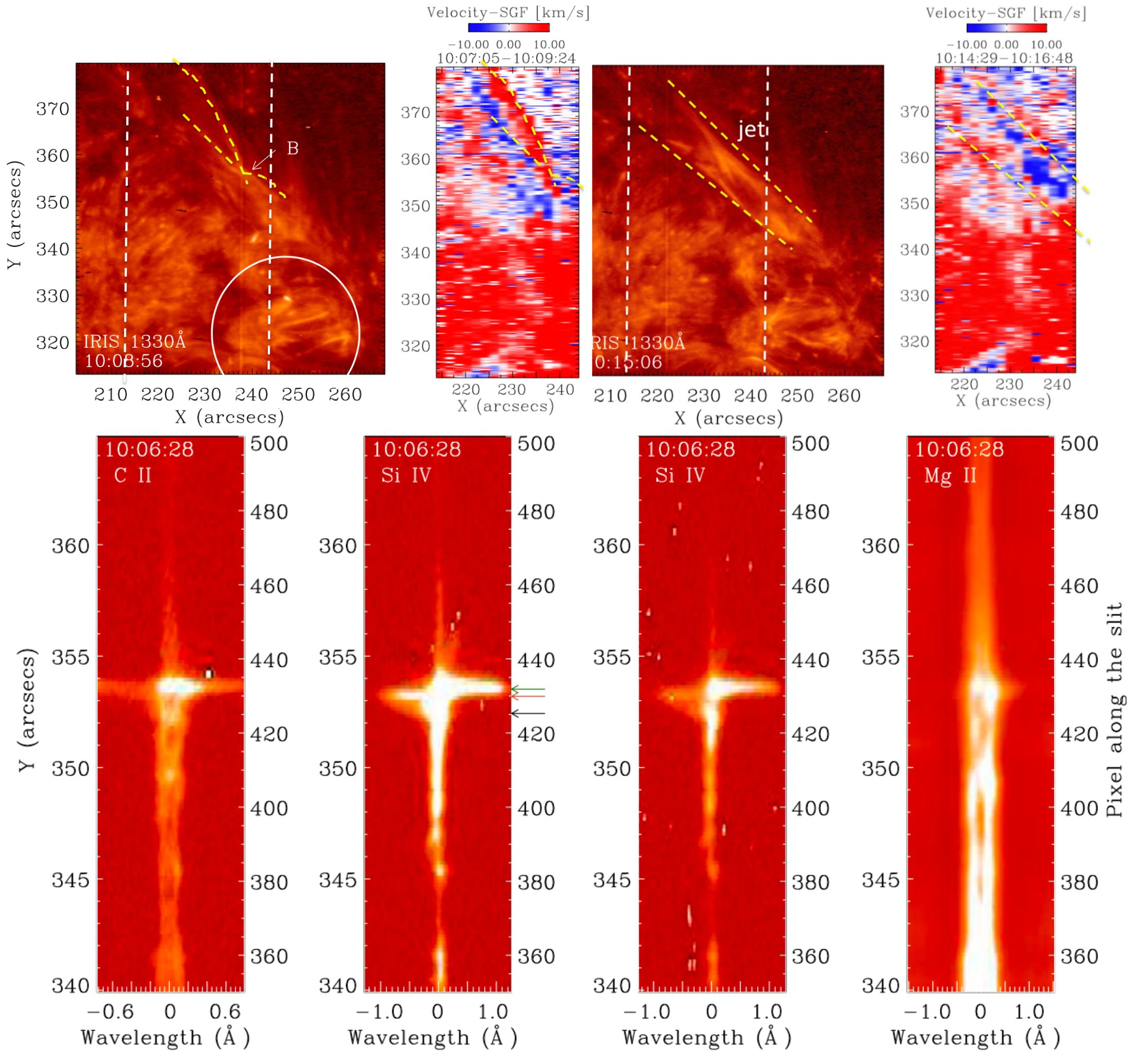}
\caption{
Jet observed with IRIS SJI and Doppler maps with blue and redshifts  showing a rotation around its main axis observed on March 30 2017 (top panels) and bidirectional flows at the reconnection point in B  in several lines (bottom panels).  The circle in the top left panel indicate the location of the emerging flux visible with arch filament system above. The vertical lines indicate the field of view of the rasters and the Doppler maps.
{The bottom panel shows the spectra of C\textsc{ii}, Si\textsc{iv} (1393.8 and 1402.8 \AA), and Mg\textsc{ii}.}
The spectra of  C\textsc{ii}, Si\textsc{iv} and Mg\textsc{ii} lines  are vertical and  C\textsc{ii} and Si \textsc{iv} spectra (but not cool lines of Mg\textsc{ii})  show an horizontal bright feature indicating high blue and redshifts around 200 km s$^{-1}$ (adapted from  \citet{Ruan2019}).}
\label{Fig_Bidirectional}
\end{figure*}

\subsection{Bidirectional flows,  rotation and twist in IRIS spectra}
\label{sec3.2}
Rotational motion is an important property of the jet-like structures in the solar atmosphere. On
the basis of Doppler velocity, it is frequently reported that  typical coronal/chromospheric
jets reveal 
%blue-shifts at its one edge while the other side plasma experiences red-shifts. 
{blue-shifts at one edge and red-shifts at the other}. It is well observed  in H$\alpha$ \citep{Ruan2019} as well in hotter lines  \citep{Pike1998} observed with the Coronal Diagnostic Spectrometer \citep[CDS,][]{Harrison1995} and  the Solar Ultraviolet Measurements of Emitted Radiation  \citep[SUMER,][]{Wilhelm1995,Curdt2011} instrument. 
Both studies used the same O\textsc{iv} transition region line, which forms at log T (K) = 5.4 and they concluded that   the spatial pattern of Dopplershifts blue on one side and red on the other side  was an evidence for helical motion. Similar  interpretation is taken for  IRIS Doppler maps  by different authors
\citep{Cheung2015,Ruan2019,Kayshap2021}. 

In \citet{Ruan2019}  a jet occurred due to  reconnection of field lines pushed upwards by an emerging flux (Fig. \ref{Fig_Bidirectional}). The reconnection point is in the corona (shown as point B) and bidirectional flows of the order of $\pm$200 km s$^{-1}$ are observed  in the spectra crossing the site, mainly in C\textsc{ii} and Si\textsc{iv} (bottom panels). The jet crossed by the IRIS slit  shows a blue and red shift pattern. Simultaneous observations in H$\alpha$ show the same blue/redshift pattern.
Such patterns suggest helical motion along the jet axis.
While \citet{Moreno2013} mention that the erupting ﬂux ropes in the simulation seem to rotate as if they were converting twist into writhe, the possible helical motions of the jets themselves were not explicitly studied.
 The conclusion of \citet{Cheung2015} concerning the
 IRIS Doppler shift maps  was that they share considerable
resemblance to synthetic Doppler maps  of \citet{Fang2014}, who carried out MHD simulations of jets resulting from the interaction of a twisting 
flux tube emerging
from the solar convection zone into a coronal layer with
ambient inclined field.

\begin{figure*}[ht!]
\centering
\includegraphics[width=0.8\textwidth]{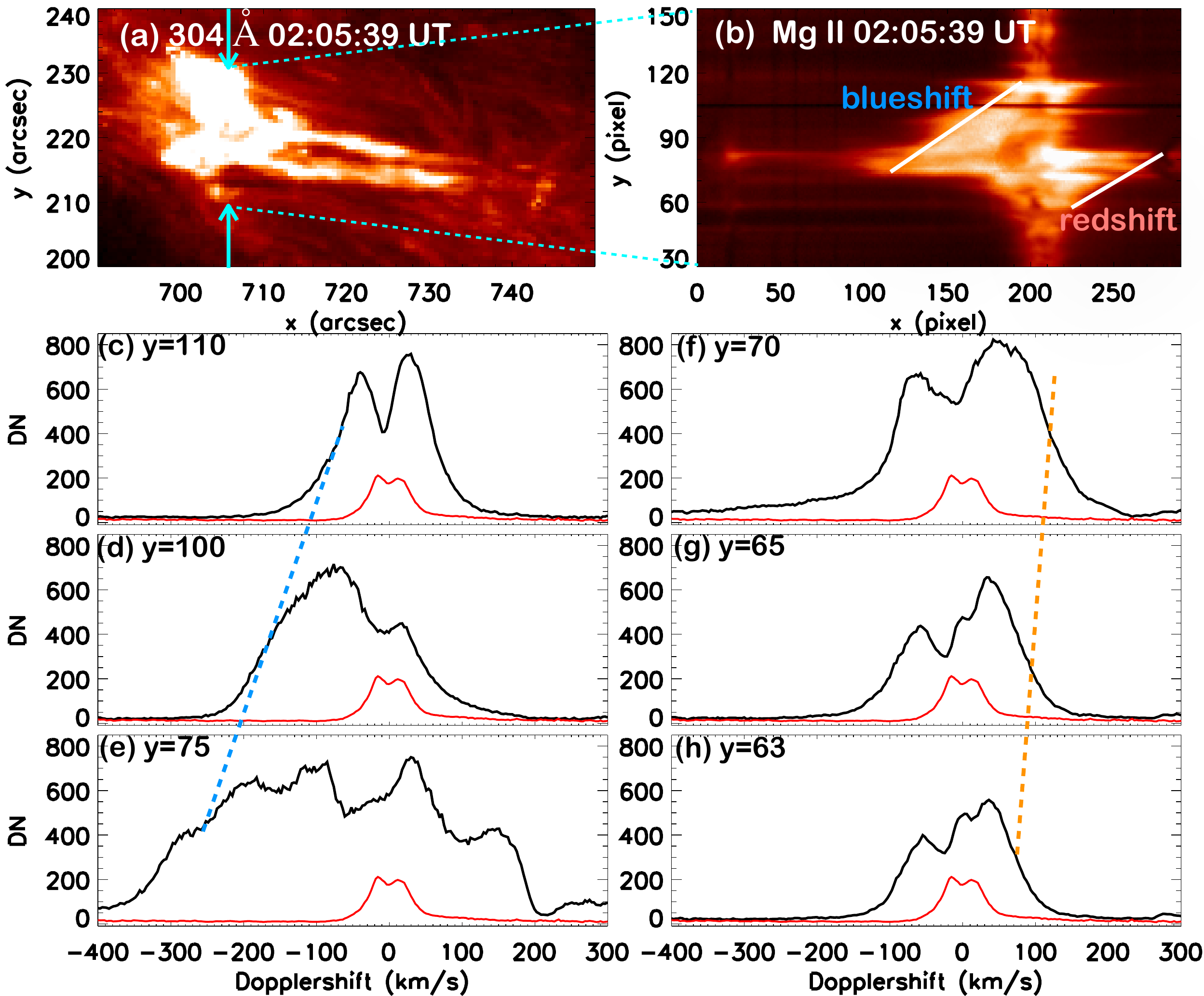}
\caption{Bright jet with two branches inserting a cool dense surge in AIA 304 Å  observed on March 22 2019 (panel a). The cyan arrows show the location of
the slit position 1. The Mg\textsc{ii}k line spectra  is crossing the reconnection site  (panel b). 
Red and blue shift wings are shown by the white tilted lines on the left (blueshift) and right (redshift) in the spectra. The bottom rows (c-h) show
the Mg\textsc{ii}k line profiles for different y values using unit of velocity in the x axis. Panels c-e concern the blue shift profiles shown by the tilted
blue dashed line corresponding to strong blueshifts {(-300  to -100 km s$^{-1}$).} Panels f-h show the red shift profiles. Redshifts (80 to 100 km s$^{-1}$) are
shown by the red dashed tilted line. The red and blue dashed lines are passing through the inflexion points of the line profiles in panels c-h). Red profiles are reference profile in each panel c to h (adapted from \citet{Joshi2020FR}}
\label{Fig_Twist}
\end{figure*}

\begin{figure*}[ht!]
\centering
\includegraphics[width=0.8\textwidth]{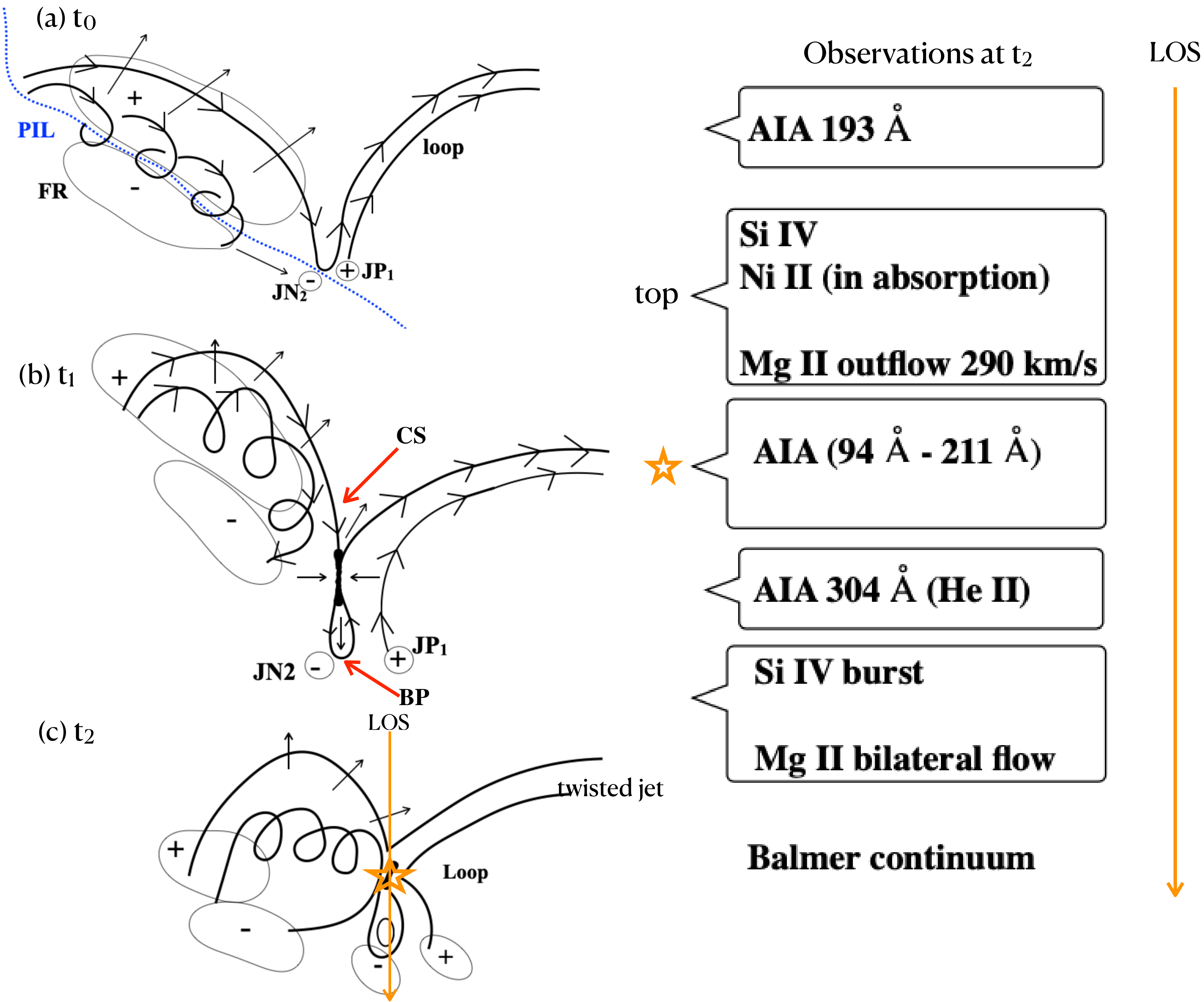}
\caption{Sketch of transfer of twist from a flux rope (FR) to a jet during reconnection in a bald patch (BP) transformed dynamically to a X-point  current sheet; ({\it left panels}):  at t$_0$  (panel a) two loop systems are approaching at the bipole JP1 and JN1, at t$_1$ (panel b)  they are tangent   at the BP forming a current sheet (CS)  and at t$_2$ (panel c) there is the reconnection in the CS (orange star);  ({\it right  panel}): flare atmosphere model along the line of sight (LOS) during t$_2$. The sketch and the flare model are deduced from the observations of IRIS, AIA and HMI vector magnetograms on March 22 2019 (adapted from \citet{Joshi2020FR}}.
\label{Fig_rot}
\end{figure*}

 In the  jet study of  \citet{Joshi2020FR}, the IRIS slit was crossing the jet base and a part of the jet (Fig. \ref{Fig_Twist} panel a).
At the pixel  of the reconnection site, the Mg \textsc{ii}
line profile is the most extended one on the blue and red sides
like in bidirectional outflows similar to the observations of \citet{Ruan2019} {and} \citet{Cai2019}.
The reconnection site Mg\textsc{ii} profile is identified  at  pixel y=75  in Fig. \ref{Fig_Twist} panel (e)).
Using a cloud model method they could interpret this wide profile by the presence of cool clouds ejected with a velocity of -300 km s$^{-1}$.  Simultaneous  Si\textsc{iv}  profiles exhibit  similar behaviour with large blueshifts.
Over these wide profiles Ni\textsc{ii} and Fe\textsc{ii} narrow lines were detected in absorption similarly as in IRIS bomb profiles \citep{Peter2014,Grubecka2016}. 
Therefore  with all the  information in the sample of IRIS spectra combined with AIA multi wavelength filtergrams they could propose  a synthesized empirical  flare model (see Fig. \ref{Fig_rot} right panel) \citep{Joshi2021b,Joshi2021a}.

In  Fig. \ref{Fig_Twist} panel (b) in the northern and southern parts of the reconnection site, the authors  note that
the spectra shows a tilt. The Mg\textsc{ii} profiles are not symmetrical
all along the slits. Therefore, with the existence
of this gradient the tilt is obvious. The tilt is characterised by
the gradient of the Dopplershifts that exist for profiles along the
jet  cross section  at a given time. The line profiles of Mg\textsc{ii}k line show important
extensions of the wings at 02:05:39 UT (Fig.\ref{Fig_Twist} panels c-h) .
The Dopplershift  velocity of the cool material
along the west side has  values of between 30 km s$^{-1}$
to 100 km s$^{-1}$ and on the east side between -300 km s$^{-1}$  to -30 km s$^{-1}$.  This means that one part of the cool material is  red-shifted
%material could be nearly normal to the solar surface
while the
other part  is {blue-shifted i.e. coming towards us.}
%would be inclined like the jet.
A similar behaviour is
observed in the four positions of the slit for Mg\textsc{ii}, C\textsc{ii}, and Si\textsc{iv} lines. The tilt in the four Si\textsc{iv} spectra is even more
readily visible because Si\textsc{iv} is a transition region line with only
one emission peak when compared to chromospheric lines with
two peaks.

The tilt in the spectra finally reach an extension  around 15 Mm, which represent the size of the cross section of the jet. The authors interpret this tilt by the
rotation of a structure at the
base of the jet or possibly cool plasma that follows helical structures. Tilt of  spectra along a slit crossing the section of the jet was first observed for
prominences \citep{Rompolt1975} and interpreted as rotating prominences
before eruption.

\subsection{Light bridge jets}
\label{sec3.3}
With  the high spatial and temporal resolution SJIs from  IRIS at 1330 \AA,  \citet{Bharti2015} 
were able to demonstrate fine scale details of jets above a light bridge (LB) in the transition zone temperature. Low altitude reconnection has been suggested as the mechanism
to produce such an activity \citep{Bharti2007}. 
\citet{Shimizu2011} reported about intermittent plasma ejections above a LB
for a few days, accompanied with a change in the photospheric magnetic flux density and
inclination. \citet{Shimizu2009} proposed a model where the LB is considered as a highly twisted
current-carrying flux tube lying below the background field that forms a cusp-like shape
above the LB. The proposed geometry then produces opposite polarity field at one side of
LB and led to reconnection. According to \citet{Toriumi2015}, the presence
of patches of high vertical current of opposite sign is indicative
of magnetic shear which is a favorable condition for
magnetic reconnection. However jets in penumbra sunspots  with LB are initiated by magneto convective process \citep{Bharti2020}.
According to the former authors current layers are formed at the edges of the convective fine structure, due to the shear between their horizontal field and the ambient vertical field. The reconnection could be  caused by an opposite polarity field produced by the bending of field lines by convective downflows at the edge of pore fine structure.
Surges have been detected at the edge of LB in  vortex formations due to the high shear existing at the edge of LB \citep{Yang2020}.  IRIS spectra would be interesting to see the Dopplershifts in these reconnection sites.
\section{Magnetic configuration in the jet environment}
\label{sec4}
\subsection{Shear between emerging flux: Transfer of twist}
\label{sec4.1}
The  jet with a strong tilt in the spectra studied in \citet{Joshi2020FR} is originated in a region where two emerging magnetic fluxes (EMFs) collapsed
 (Fig. \ref{Fig_rot} left panel).
Magnetic vector maps exhibit a long sigmoidal flux rope (FR) 
along the polarity inversion line between the two EMFs, which is the site
of the reconnection. Before the jet, an extension of the FR was present and a part of it was detached and formed a small bipole with a
bald patch (BP) region, which dynamically became an X-current sheet over the dome of the eastern  EMF where the reconnection took place.  This view is consistent with simulations  of IRIS bombs proposed by \citet{Hansteen2019}.

At the time of the reconnection, the Mg\textsc{ii} spectra exhibited a strong tilt,  this was the signature of the transfer of the twist to the jet (Fig. \ref{Fig_Twist}).
 A comparison  of the vector magnetic   and electric current maps with numerical magnetohydrodynamics (MHD) simulations \citep{Janvier2013,Aulanier2012,Aulanier2019} confirms the existence of the long FR. The authors
conjectured that there was a transfer of twist to the jet during the extension of the FR to the reconnection site without FR eruption. The
reconnection would start in the low atmosphere in the BP reconnection region and extend at an X-point along the current sheet formed
above like in the 3D  MHD simulations of \citet{Hansteen2019} and  \citet{Wyper2019}.

For this event,
\citet{Joshi2020FR} could propose a multi thermal atmosphere model of the mini flare at the jet base  as it was mentioned in the previous section  (Fig. \ref{Fig_rot} right panel).  The reconnection was accompanied by a bombardment of energetic electrons detected by hard X-ray emission measured by the  FERMI/Gamma Burst Monitor \citep[GBM,][]{Meegan2009} 
\citep{Joshi2021b}. These energetic particles produced an enhancement of the Balmer continuum observed by IRIS. The good timing was the main favorable argument of this relationship.  This phenomena  is commonly observed for {GOES X-class  flare} \citep{Heinzel2014,Kleint2016,Kleint2017}.  Therefore \citet{Joshi2021b} concluded that the energetic electrons  could be efficient for {a B class flare} only if  reconnection occurred in a tiny area smaller than the IRIS spatial resolution.

\subsection{Emerging flux models}
\label{sec4.2}
In \citet{Joshi2020}
the series of jets and surges provides a good case study for testing the 2D and 3D magnetohydrodynamic emerging
flux models  even the jets were observed at the limb \citep{Nobrega2017} (Fig. \ref{Fig_Nobrega} panel e). The double-chambered structure that is found in the observations corresponds to the regions with cold and hot loops that
are in the models below the current sheet that contains the reconnection site. The cool surge with kernels is comparable with the cool
ejection and plasmoids that naturally appears in   current sheet models \citep{Ni2016,Ni2021}.

\subsection{Current sheet formation initiating jets}
\label{sec4.3}
Accumulation of electric currents observed in photospheric magnetic vector maps is a good indicator for
predicting flares, eruption and jets. Photospheric motions of magnetic polarities lead to  stressed  magnetic field lines which can release energy during reconnection.  
Photospheric electric current pattern evolution was observed of the jet bases \citep{Guo2013}. The authors    found  in the photospheric  magnetic topology analysis that the pattern was  associated with  quasi-separatrix layers (QSL)  deduced from  magnetic extrapolation.  There was a strong correlation between the built up of electric currents along the QSLs and the flux evolution of the jet base.

The built up of electric current is directly tied to  photospheric motions. An other case  on March 22, 2019 
where   IRIS jets were strongly related to  photospheric motions which built up strong electric currents detected in HMI vector magnetic field maps \citep{Joshi2020FR}.
In that  paper  it was  also suggested that electric currents were storage in the remnant {FR} magnetic field lines overlying the emerging dome. The transfer of twist occurred between the FR and the jet during reconnection.

\subsection{Braiding loops and nano jets}
\label{sec4.4}
Bidirectional  flows were detected during the reconnection in the IRIS spectra as Dopplershifts \citep{Ruan2019,Joshi2020FR}. Therefore the  direction of bidirectional ejections could be perpendicular  to the apparent transverse flows of the jets if measured in the sky plane. The bidirectional flows come from the  reconnection-jet  at the reconnection site  (a few pixels in IRIS spectra).  
This phenomena is similar to the nano jets observed during the braiding of magnetic field lines of a loop attached to a filament \citep{Antolin2021}. The change of curvature of the field lines leads to the release of electric currents stored in the field lines and nano jets were detected in the IRIS spectra of transition zone plasma with speed of 200 \kms. {Recently a statistical analysis of 43 nano jets observed with IRIS was done by \citet{Chen2020} where the jets were  explained   by  fine-scale external/internal magnetic reconnection.} \citet{Antolin2021} used MHD simulations to 
demonstrate that  nano jets could be  a consequence of the slingshot effect from the magnetically tensed, curved magnetic field lines reconnecting at small angles and produced coronal heating plasma at the reconnection site. 
Such release of energy could  relatively be frequent in the corona as predicted {by \citet{Parker1988} and \citet{Demoulin1997}} and observed recently with Solar Orbiter, named as campfires \citep{Berghmans2021}.
Dissipation of magnetic energy in the corona requires the creation of very fine scale-lengths because of the high magnetic Reynolds number of the plasma. The formation of current sheets is a natural possible solution to the heating problem and it is  known that  magnetic field  stressed by continuous photospheric motions through a series of equilibria can easily form such sheets \citep{Demoulin1997}.

\section{Summary and Conclusion}
\label{sec5}
Recent  jet observations with IRIS, spectra and SJIs, {benefit from the high spatial} resolution of the instrument (pixel size = 0.167 arcsec) and the high temporal cadence, never reached for  chromosphere and  transition region  temperature spectra.
{These observations revealed that solar jets are very dynamic} and  have an impulsive nature like  explosive events. {Jet base is heated over a large range of temperatures. The reconnection can be accompanied by energetic electron ejections like in X-ray flares. 
Bidirectional flows  are detected in reconnection sites, they correspond to jet-reconnections} 
\citep{Ruan2019,Joshi2021a,Antolin2021}.

{The main difference between jets and flares or eruptions is the size of the reconnection site. In the case of jets, the reconnection region is very small and the way how reconnection occurs is thus difficult to observe.}
{Currently, it is impossible to resolve magnetic configurations at the reconnection site in the environment of jets.} It is only by  MHD simulations that we may assume the reconnection mechanism.

We show the importance of 
 the role of photospheric motions and some attempts have been made to  
quantify the  electric currents   and related directly to 
recurrent jets 
\citep{Guo2013}. During 
  emerging magnetic flux regions or filaments, the role of stress and accumulation of electric currents is important 
\citep{Joshi2020FR}.
{However,} many questions are still open, the transfer of twist, the role of the convection in the light bridge jets, the importance of the braiding of  magnetic field lines to answer to  coronal heating, the  role of network  jets in  the corona as sources of solar wind.
Solar Orbiter and Parker Solar Probe already brought some answers with their discoveries of camp-fires \citep{Berghmans2021} and switchbacks of magnetic field \citep{Bale2019,Ruffolo2020}.

\section{Acknowledgments}
IRIS is a NASA small explorer mission developed and operated by LMSAL with mission
operations executed at NASA Ames Research center and major contributions to downlink
communications funded by the Norwegian Space Center (NSC, Norway) through an ESA
PRODEX contract. We thank the SDO/AIA, SDO/HMI, and IRIS science teams for granting free access to the data. 
{We thank the reviewers for their valuable suggestions which were quite important for us to improve the paper.}
RJ thanks to CEFIPRA for the Indo-French Raman Charpak fellowship. RJ and RC acknowledge the support from Bulgarian Science Fund under Indo-Bulgarian bilateral project, DST/INT/BLR/P-11/2019. This paper was presented at the COSPAR 2020 General Assembly session E2.2 in February 2021.

%\bibliographystyle{model5-names}
%\bibliography{refs}
\bibliography{COSPAR2021_ASR_Revised_Clean.bbl}

\end{document}